\documentclass[letter]{aa}  

\usepackage{graphicx}
\usepackage{txfonts}
\definecolor{light}{rgb}{0.6,0.6,0.7}

\newcommand{\Msun  }{M_\odot}
\newcommand{\Rsun  }{R_\odot}

\graphicspath{{./}{plots/}}

\begin{document} 

\title{Planet--planet scattering as the source of the highest eccentricity exoplanets}

\author{
Daniel Carrera\inst{1}\thanks{E-mail: dcarrera@gmail.org}\and
Sean N. Raymond\inst{2}\and
Melvyn B. Davies\inst{3}
}

\institute{
Department of Astronomy and Astrophysics, 525 Davey Laboratory, The Pennsylvania State University, University Park,\\PA 16802, USA\and
Laboratoire d$'$astrophysique de Bordeaux, Univ. Bordeaux, CNRS, B18N, all\'ee Geoffroy Saint-Hilaire, 33615 Pessac, France\and
Lund Observatory, Department of Astronomy and Theoretical Physics, Lund University, Box 43, SE-221 00 Lund, Sweden
}

\date{Received XXX; Accepted YYY}

\abstract{
%
%
Most giant exoplanets discovered by radial velocity surveys have much higher eccentricities than those in the solar system. The planet--planet scattering mechanism has been shown to match the broad eccentricity distribution, but the highest-eccentricity planets are often attributed to Kozai-Lidov oscillations induced by a stellar companion.
%
%
Here we investigate whether the highly eccentric exoplanet population can be produced entirely by scattering.
%
%
We ran 500 N-body simulations of closely packed giant-planet systems that became unstable under their own mutual perturbations.
%
%
We find that the surviving bound planets can have eccentricities up to $e > 0.99$, with a maximum of 0.999017 in our simulations. This suggests that there is no maximum eccentricity that can be produced by planet--planet scattering.
Importantly, we find that extreme eccentricities are not extremely rare; the eccentricity distribution for all giant exoplanets with $e > 0.3$ is consistent with all planets concerned being generated by scattering.
%
%
Our results show that the discovery of planets with extremely high eccentricities does not necessarily signal the action of the Kozai-Lidov mechanism.
}

\maketitle
%
%
\section{Introduction}

One of the most interesting discoveries over the short history of exoplanet science is that although planetary systems are common in the Galaxy, most of them are in many ways quite different from the solar system \citep[see, e.g.][]{Raymond_2018}. One of those differences is that many planets have  a much more eccentric orbit than that of any planet in the solar system \citep[e.g.][]{Udry_2007}, with the most massive exoplanets having the most eccentric orbits \citep{Jones_2006,Ribas_2007,Ford_2008,Raymond_2010}.

Gravitational interactions between planets in a system lead to a chaotic evolution of orbital elements that can cause planet orbits to cross. When that happens, giant planets rarely collide, but instead gravitationally scatter \citep{Rasio_1996,Weidenschilling_1996,Lin_1997}; for a review, see \citet{Davies_2014}. Planet--planet scatterings can reproduce much of the observed distribution of exoplanet eccentricities for a wide range of initial conditions \citep{Adams_2003,Moorhead_2005,Chatterjee_2008,Ford_2008,Juric_2008,Raymond_2010,Raymond_2011}. See also \citet{Davies_2014} for a review of planet-planet scattering. A planetary system is said to be \textit{Hill stable} if the orbits of the planets therein will never cross. For a two-planet system it is possible to derive, to first order, a criterion that guarantees that the system is Hill stable \citep{Zare_1977,Marchal_1982,Gladman_1993}. For systems with three or more planets, our understanding of the stability limits comes largely from numerical experiments. For example, \citet{Chambers_1996} showed that systems with three equal-mass planets ($m_p / M_\star \in \{10^{-5}, 10^{-7}, 10^{-9}\}$) are probably always unstable if $\Delta < 10$, where $\Delta$ is the planet separation in terms of their mutual Hill radii,

\begin{eqnarray}\label{eqn:Delta}
        \Delta &=& \frac{a_2 - a_1}{R_{\rm Hill}}, \\
        R_{\rm Hill} &=&
                    \left( \frac{m_1 + m_2}{3 M_\star} \right)^{1/3}
                    \left( \frac{a_1 + a_2}{2} \right),
\end{eqnarray}
where $m_1$, $m_2$, $a_1$, and $a_2$ are the masses and semimajor axes of the two planets, and $M_\star$ is the stellar mass. Similar experiments have been conducted by several authors \citep[e.g.][]{Marzari_2002,Chatterjee_2008,Marzari_2014}. Broadly speaking, the evolution of the planetary system is highly chaotic and there are various islands of stability \citep{Marzari_2014}, but in general, the time to close encounters increases rapidly with $\Delta$ and decreases with the number of planets in the system \citep{Chambers_1996}. Close encounters then lead to strong dynamical scatterings and sudden changes in the orbital parameters. To avoid most of the dependence on the number of planets, \citet{Faber_2007} ran simulations for ten-planet systems and estimated that

\begin{equation}
    \log_{10}\left(\frac{t_{\rm ce}}{\rm yr}\right)
    =
    -1 - \log_{10}\left(\frac{\mu}{10^{-7}}\right)
    + 2.6\,\Delta\,\mu^{1/12},
\end{equation}
where $t_{\rm ce}$ is the time to the first close encounter and $\mu$ is the planet--star mass ratio. More recently, \citet{Marzari_2014} showed that the stability structure of a three-planet system is a two-dimensional space. Rather than ``stability islands'', the stable region forms a grid structure caused by the mean-motion resonances between adjacent planet pairs. It is important to keep this structure in mind when studying the stability of three-planet systems.

The orbit-crossing phase lasts until at least one planet is removed from the system, either by a collision with another body, or by being ejected from the system. The planets that survive are left with higher eccentricities. Planetary systems with equal-mass giant planets tend to produce the highest eccentricities \citep{Ford_2008,Carrera_2016}. To reproduce the overall eccentricity distribution of giant exoplanets probably requires a mix of equal-mass and unequal-mass systems \citep[][Figure 13]{Carrera_2016} with the most massive planets typically forming equal-mass systems \citep{Raymond_2010,Raymond_2012,Ida_2013}. In this investigation we are interested in the highest eccentricities, so we focus on equal-mass systems.

\citet{Ford_2008} suggested that the maximum eccentricity that can be attained from planet--planet scatterings, regardless of the planet mass ratio, is around $e \sim 0.8$. Importantly, that conclusion is conditioned on the two-planet scenario considered by the authors. Other simulations involving three or more giant planets have been seen to produce eccentricities above $0.9$ \citep[e.g.][]{Veras_2006,Chatterjee_2008,Nagasawa_2008,Veras_2009,Zanardi_2017}, but a targeted investigation of the limits of planet--planet scattering has so far been absent in the literature. The significance of this is that if there is indeed an ``$e_{\rm max}$'' from scattering, that would mean that any planet with $e > e_{\rm max}$ must have acquired its eccentricity through some other process, such as the Kozai-Lidov effect \citep{Kozai_1962,Lidov_1962}. In other words, if past experiments have lead us to an ``$e_{\rm max}$'' that does not exist, we may be misinterpreting the evidence from exoplanet surveys, and drawing incorrect conclusions about which mechanisms have shaped the exoplanet population.

In this work we probe the limits of planet--planet scattering, both in terms of the maximum eccentricity that can be achieved, and how often $e \sim 1$ could be plausibly produced. It is no use to say that planet--planet scattering can produce an eccentricity of $\sim 1$ if those events are so rare that they cannot be part of the observed exoplanet population.

This Letter is organised as follows. In Section \ref{sec:methods} we describe our simulations and initial conditions. In Section \ref{sec:results} we present our results. We discuss our results in Section \ref{sec:discussion} and draw conclusions in Section \ref{sec:conclusions}.

%
%
\section{Methods}
\label{sec:methods}

We ran 500 N-body simulations using the \textsc{mercury} code with the hybrid integrator \citep{Chambers_1999}. All simulations had a single star with a mass of $1 \Msun$, and three Jupiter-like planets, each with a mass of $10^{-3} \Msun$  and $\rho = 1.4$ g cm$^{-3}$ (about the same density as Jupiter). We choose equal-mass planets in order to maximise the final eccentricities after the dynamical instability \citep{Carrera_2016}. In addition, \citet{Raymond_2010} proposed that giant planets are more likely to be born in equal-mass systems.

Table \ref{tab:init} shows the initial orbital parameters of all the planets. The innermost planet was placed at 3 AU, and the other planets were arranged so that the planets were all separated by five mutual Hill radii, meaning that the simulated systems become unstable quickly. While most giant planets from RV surveys have semimajor axes less than 3 AU, placing the planets farther from the star allows them to be more eccentric without colliding with the star or becoming tidally circularised. We inflated the stellar radius to 0.03 AU (roughly $6 \Rsun$) because $q = 0.03$ AU seems to be the empirical limit for tidal circularisation \citep[e.g.][Figure 4]{Beauge_2012}. In other words, planets that would normally have their orbits circularised by stellar tides will instead collide with our simulated star. For our numerical experiment to produce believable results, it is essential to accurately resolve close passages with the central star, which can be a source of energy error if the time-step is too large \citep{Rauch_1999,Levison_2000,Raymond_2011}. In order to minimise numerical errors, we used an integration time-step of 0.1 days, and an accuracy parameter of $10^{-11}$. We set the ejection radius to $10^5$ AU, so as to not preemptively remove planets.

\begin{table}
  \caption{Initial conditions. We ran 500 N-body simulations. Each simulation had three Jupiter-like planets separated by five mutual Hill radii. The eccentricity and inclination were fixed, and the other orbital angles (longitude of periastron, longitude of ascending node, mean longitude) were randomised.}
  \label{tab:init}
  \begin{tabular}{lcccc}
  Planet & $a$ (AU) & $e$ & $I$ (deg) & $\omega,\Omega,\lambda$ (deg)\\
  \hline
  \tt{J1} & 3.0000 & 0.05 & 1$^\circ$ & random $\sim$ U[0-360$^\circ$) \\
  \tt{J2} & 4.6765 & 0.05 & 1$^\circ$ & random $\sim$ U[0-360$^\circ$) \\
  \tt{J3} & 7.2899 & 0.05 & 1$^\circ$ & random $\sim$ U[0-360$^\circ$) \\
  \end{tabular}
\end{table}

All planets started out in near circular ($e = 0.05$), near coplanar orbits ($I = 1^\circ$ in the lab frame), and all other orbital angles were randomised. Therefore, mutual inclinations range from $0^\circ$ to $2^\circ$, with a median of $1.4^\circ$. We ran the simulations for 10 Myr. At the end of the simulations, we determined the final eccentricities and semimajor axes of the surviving planets. We removed any planets that were in hyperbolic orbits, as they are in the process of escaping the system.

Finally, in order to compare our simulations against observations, we downloaded the exoplanet catalogue from \url{exoplanets.org}\footnote{Downloaded on February 5, 2019}. We selected all the planets discovered by radial velocity with a measured mass of $m \sin(I) > 1 M_{\rm Jup}$.

%
%
\section{Results}
\label{sec:results}

\begin{figure}
        \includegraphics[width=\columnwidth]{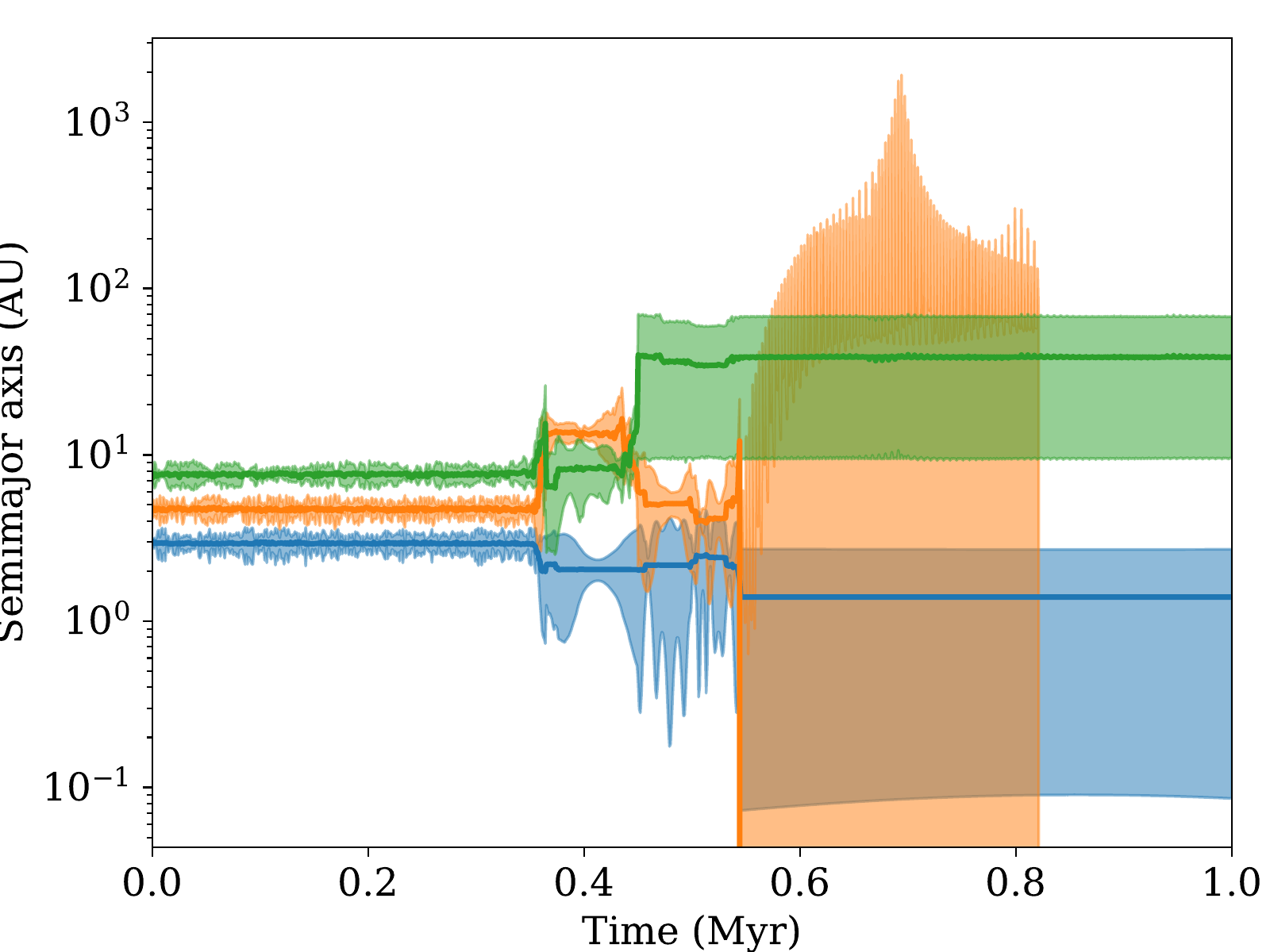}
    \caption{Sample simulation. The solid lines mark the semimajor axis of each planet, and the shaded region goes from periastron to apastron. The three Jupiter-like planets start in near coplanar, near circular orbits. After a period of chaotic evolution but mostly stable orbits, there is a sudden instability. This leads to a period of orbit crossing and strong interactions that only ends when one of the planets is ejected from the system at $t = 0.8$ Myr. The remaining planets have final eccentricities of $e = 0.754$ (outer planet) and $e = 0.938$ (inner planet) at 10 Myr.}
    \label{fig:example}
\end{figure}

Figure \ref{fig:example} shows the evolution of one of our simulations. A typical instability begins with a period where the orbital parameters evolve chaotically, but orbits remain separated. There is then a sudden instability that leads to a period of orbit crossing and associated close encounters between the planets. This period ends with the removal of one of the planets from the system. In our simulations, 311 of 500 runs had a planet ejected, 113 had a collision with the host star, and 109 had a planet--planet collision (some systems had more than one of these events). In the end, 82 runs ended with one giant planet, 392 had two giant planets, and 26 runs still had three giant planets. The run with the largest error had $|\Delta E / E| = 2.967 \times 10^{-3}$. The median energy error was $|\Delta E / E| = 2.858 \times 10^{-7}$ and 95\% of the runs had $|\Delta E / E| < 10^{-3}$. The errors in angular momentum were smaller, with a maximum $|\Delta L / L|$ of $1.503 \times 10^{-4}$. See Appendix \ref{sec:appendix} for an in-depth look at the integration errors.

\begin{figure}
        \includegraphics[width=\columnwidth]{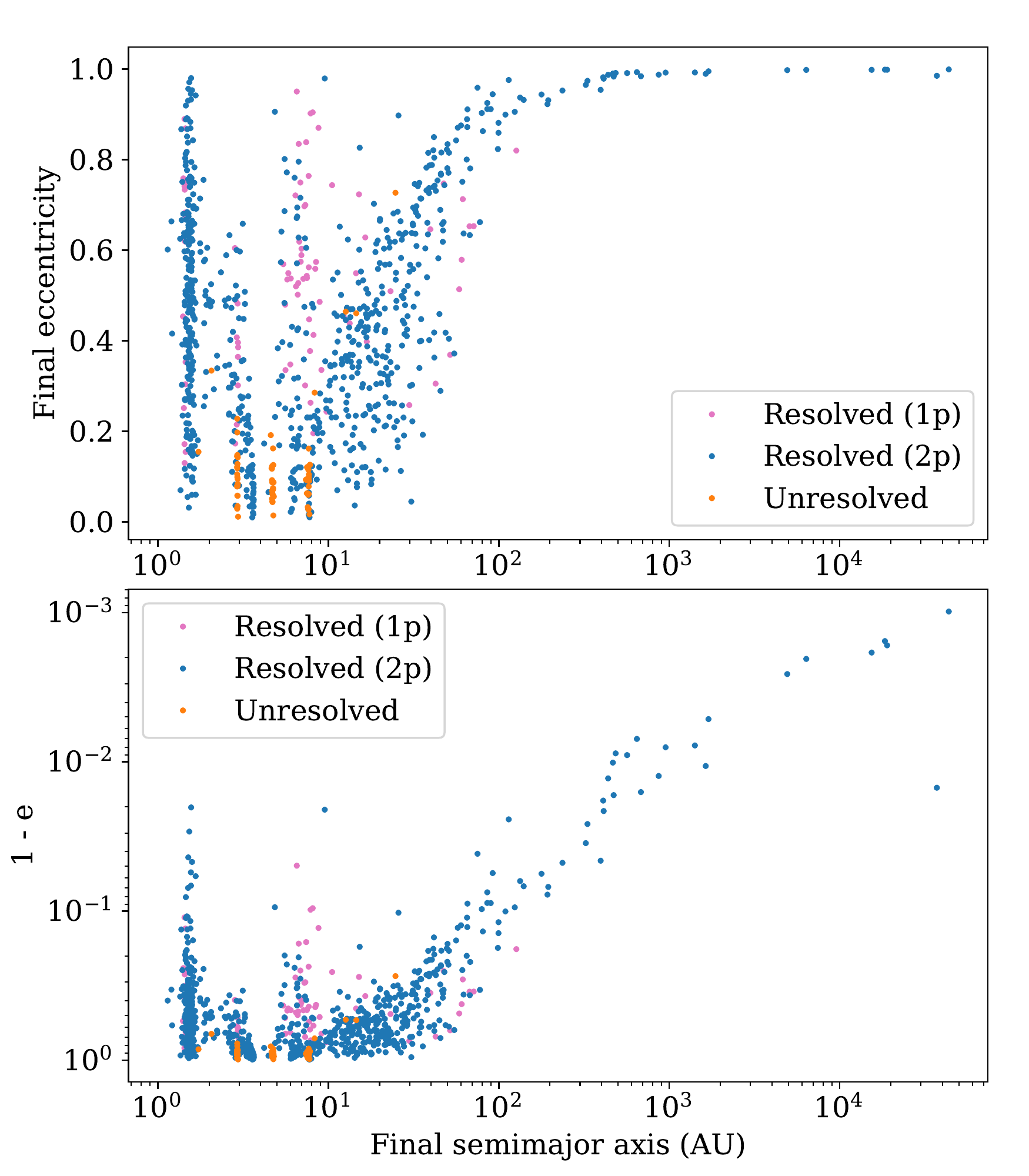}
    \caption{Final semimajor axes and eccentricities of all the simulated giant planets that survived to the end of the 10 Myr simulation. We say that a simulation is resolved if there was at least one ejection or collision. Most runs result in two surviving planets (blue) and nearly all planets with $e > 0.9$ appear in two-planet systems. For unresolved systems, some of the planets may be ejected in the future.  Unresolved systems are excluded from our analysis to avoid this potential source of bias.}
    \label{fig:ecc}
\end{figure}

Most of our systems have their first close encounter very quickly. The median time to the first close encounter was only $t_{\rm ce} = $1,885 yr, and 
95\% of the systems had $t_{\rm ce} < $ 40,000 yr. There was a wide spread in time between the first close encounter and the first ejection or collision, with a central 90\% interval of 35,000 yr to 4.6 Myr.

Figure \ref{fig:ecc} shows the final eccentricities and semimajor axes of all the giant planets that survived to the end of the simulation. There were 26 systems that still had all three giant planets at the end of the 10 Myr integration. Three of those had a planet on a hyperbolic orbit, and were therefore treated as an ejection. We consider the other systems ``unresolved'', since it is likely that some of those systems will have collisions or ejections at some point in the future. Resolved systems almost always end up with two giant planets, and nearly all planets with $e > 0.9$ belong to a two-planet system. Among the systems that are resolved, the most eccentric planet had a final eccentricity of $e = 0.999017$. In other words, our study suggests that there is no maximum eccentricity that can be produced by planet--planet scattering events. To illustrate why our result was not discovered earlier, out of the 33 resolved planets with $e > 0.95$, 24 have semimajor axis beyond 300 AU and 29 had an apastron above 100 AU, which would have been considered an ejection by many previous sets of simulations \citep[e.g.][]{Raymond_2010,Raymond_2011}. The other four planets {all} had periastrons that at some point were beyond the numerical resolution of \citet{Raymond_2010,Raymond_2011}.

\begin{figure}
        \includegraphics[width=\columnwidth]{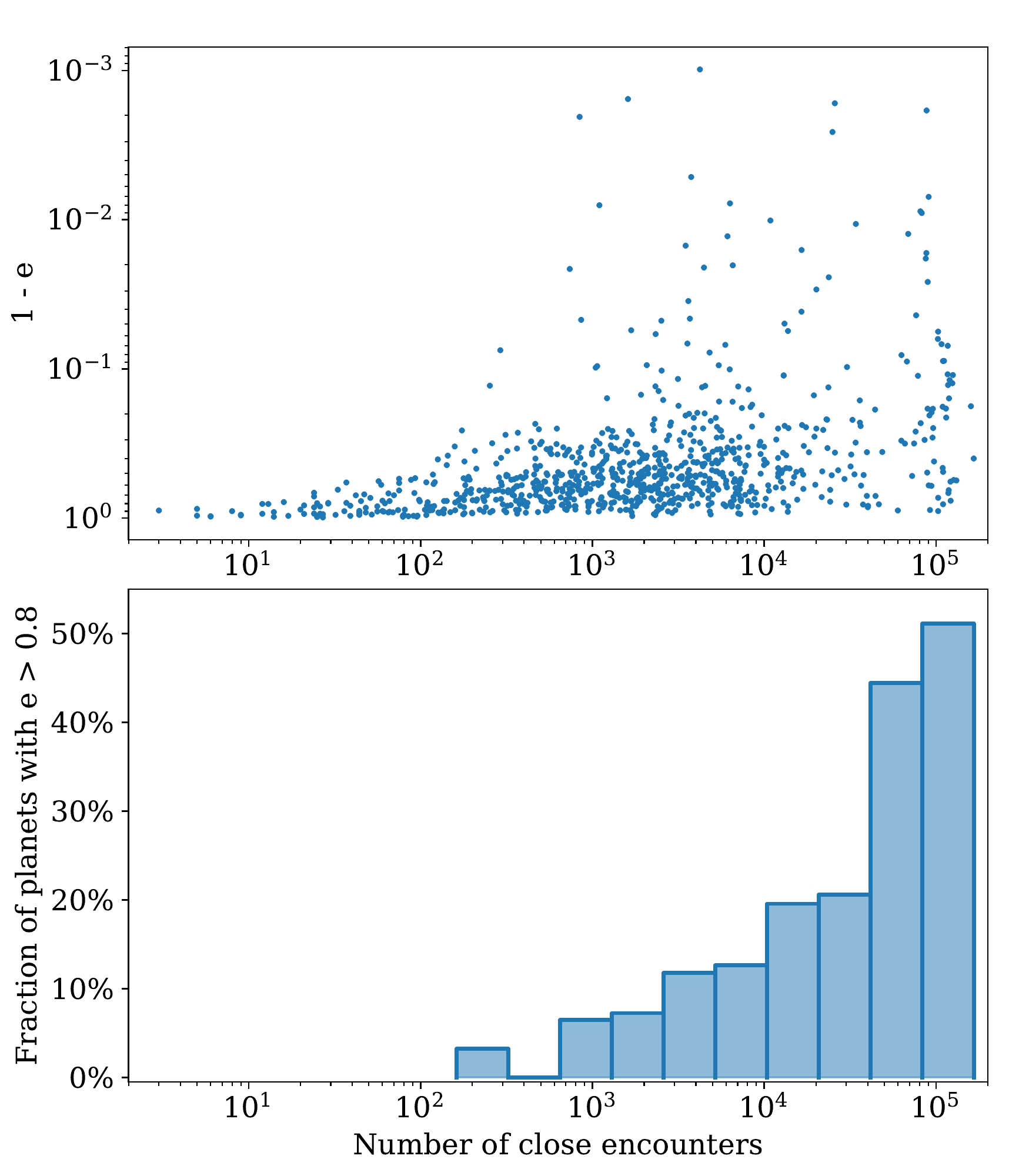}
    \caption{Final eccentricities of all ``resolved'' planets vs. the number of close encounters that each planet experienced. A close encounter is defined as two planets having a closest approach inside 3 Hill radii. The greater the number of close encounters, the greater the probability that the giant planet will reach an extremely high eccentricity.}
    \label{fig:ecc-vs-Nce}
\end{figure}

\begin{figure}
        \includegraphics[width=\columnwidth]{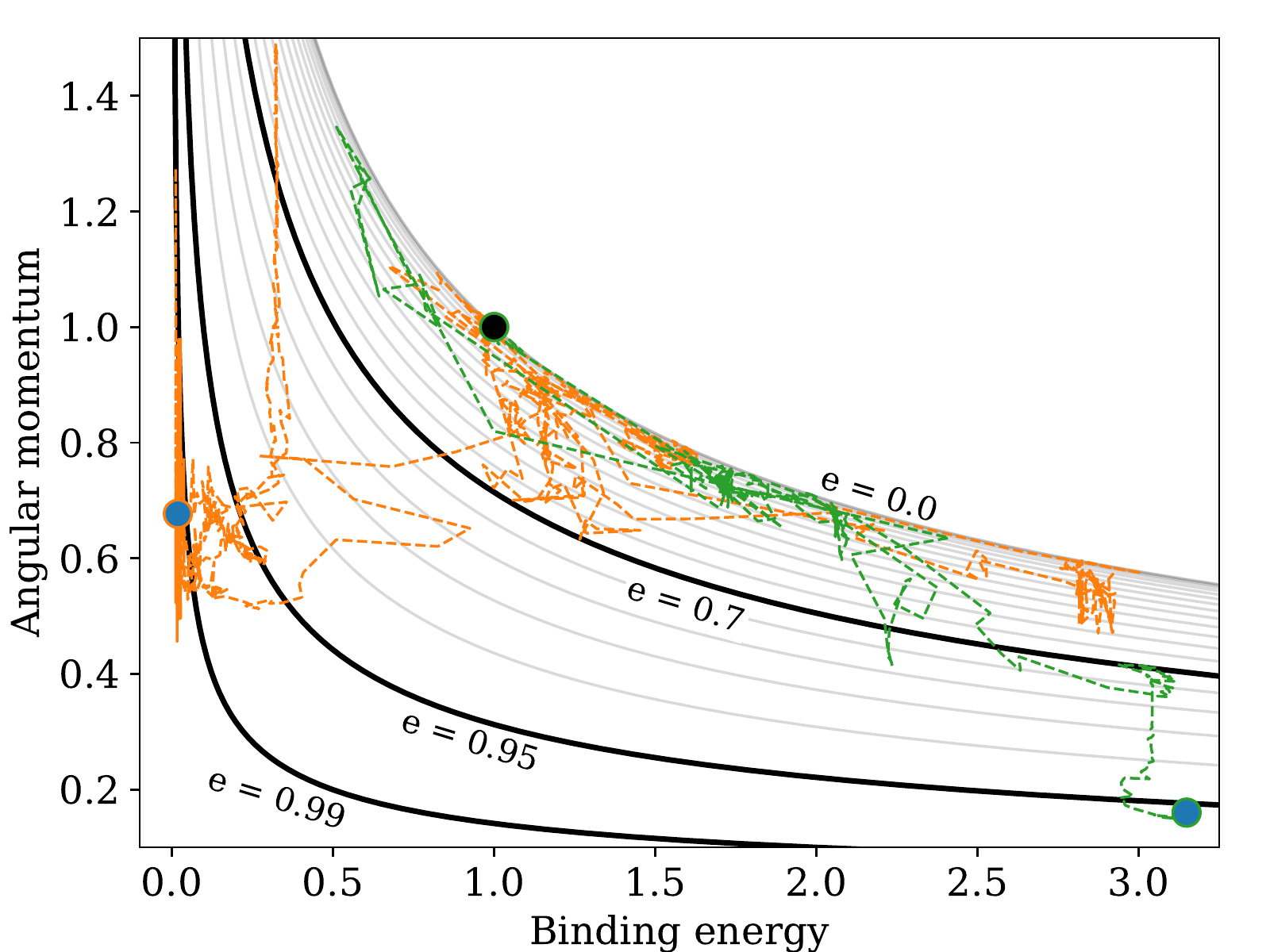}
    \caption{Evolution of the binding energy and angular momentum of two planets that reached $e > 0.95$. Values are normalised to the initial orbit of the planet (black dot). We also show constant-eccentricity curves spaced in intervals of $\Delta e = 0.05$, except we show $e = 0.99$ instead of $e = 1$. The orange trajectory follows a planet whose semimajor axis increased from 7.3 to 395 AU, while its angular momentum dropped by only 30\%. The green trajectory is for a planet that in fact \textit{gained} binding energy, but lost 84\% of its angular momentum. For clarity, only the first 1 Myr is shown. The final eccentricities shown are 0.996 and 0.959, respectively.}
    \label{fig:random-walk}
\end{figure}

The dynamical pathway that produces these extremely eccentric planets is characterised by a large number of close encounters. Figure \ref{fig:ecc-vs-Nce} shows planet eccentricities against the number of close encounters (i.e. encounters within 3 $R_{\rm Hill}$) that the planet experienced. Figure \ref{fig:random-walk} shows examples of the two types of dynamical histories that can lead a planet to an extreme eccentricity. For a planet to reach $e > 0.95$, it needs to either gain a great deal of energy (orange path; near total loss of binding energy), or it needs to lose most of its angular momentum (green path). A sequence of close encounters causes the orbit of the planet to follow a random walk across the energy-momentum phase space. The more encounters, the greater the chance that the planet will reach the high-eccentricity region of the phase space.

The most eccentric planets ($e > 0.99$) inevitably have large semimajor axes (i.e. their dynamical histories resemble the orange trajectory in Figure \ref{fig:random-walk}). For example, the one planet with $e = 0.999017$ also had $a = 43,656$ AU. This is inevitable if the planet is to avoid stellar collisions or tidal circularisation. A planet with $e = 0.995$ at 3 AU would have its periastron at $q = 0.03$ AU and would be at risk of being tidally circularised. In order to make a more direct comparison with observations we select the simulated planets with semimajor axes less than 5 AU. The most eccentric planet with $a < 5$ AU in a ``resolved'' system has $e = 0.98$. This value is just above the highest measured eccentricity for an exoplanet, which is $e = 0.97$ \citep[HD 20782 b,][]{OToole_2009}.

\begin{figure}
        \includegraphics[width=\columnwidth]{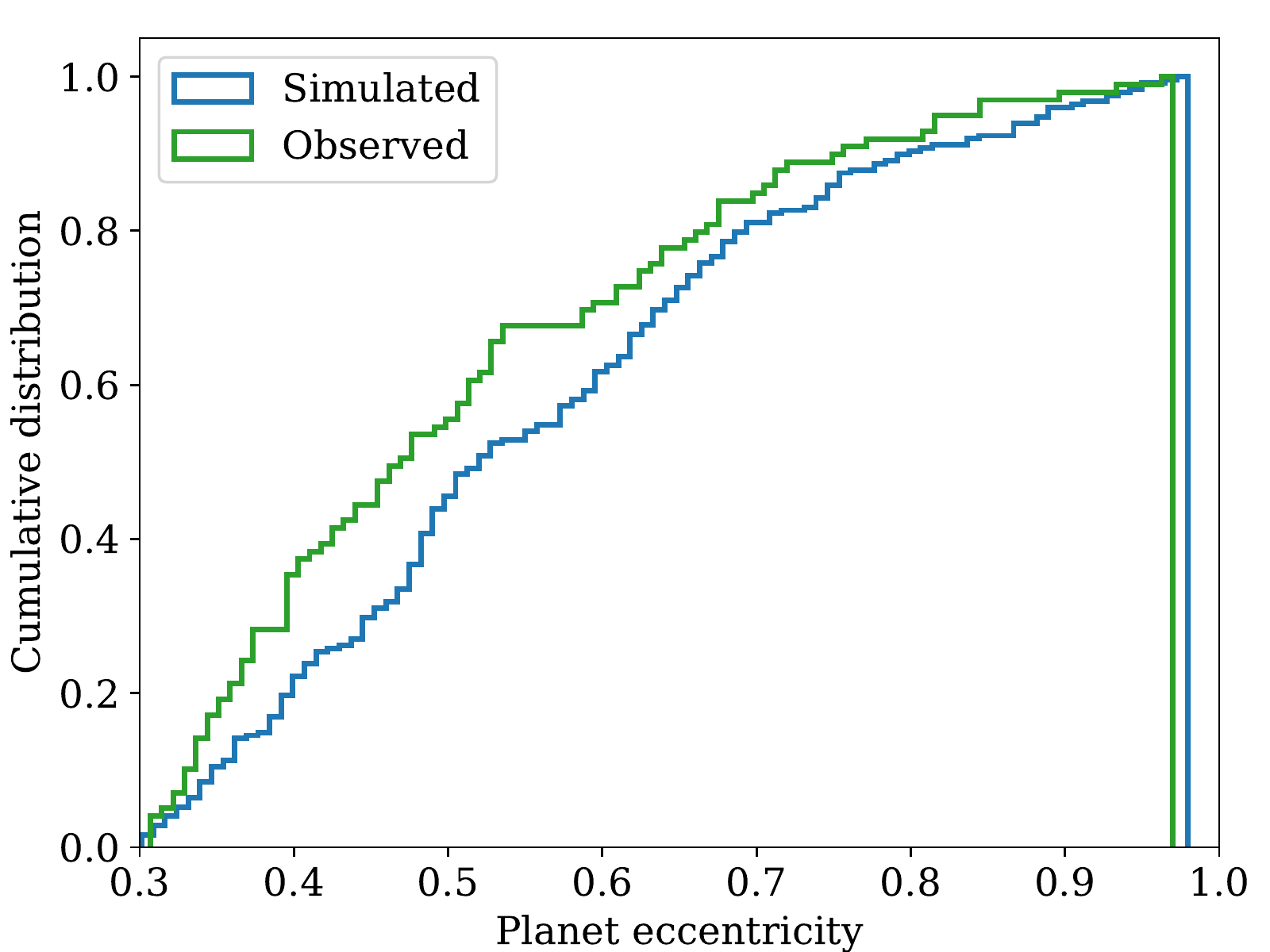}
    \caption{Cumulative distribution of eccentricities for observed and simulated giant planets with $e > 0.3$. The simulated sample is restricted to planets in ``resolved'' systems (see main text) with semimajor axis below 5 AU. The observed sample is restricted to planets with $m \sin(I) > 1 M_{\rm Jup}$.}
    \label{fig:sim_vs_obs}
\end{figure}

Figure \ref{fig:sim_vs_obs} shows the cumulative distribution of eccentricities for simulated and observed giant planets with $e > 0.3$. The figure shows no evidence of an excess of highly eccentric planets in the RV sample. Quite the contrary, the simulated planets are slightly more eccentric. This would be expected if $e > 0.3$ eccentricities come from scattering, because giant exoplanets probably do not always form in systems with {exactly} equal masses. While this result does not imply that the Kozai-Lidov effect never occurs, it clearly suggests that Kozai-Lidov is not a major force in shaping the eccentricity distribution of giant exoplanets (or it might imply that the Kozai-Lidov mechanism coincidentally produces an eccentricity distribution similar to that from planet scatterings; but we believe that option to be unlikely).

Finally, it is worth commenting on the fact that Figure \ref{fig:sim_vs_obs} compares present-day observed systems which are typically a few gigayears old with simulations that only lasted 10 Myr. The reason why sub-gigayear simulations  are valid and commonly used is that once planets have a dynamical instability, the survivors cluster just past the edge of the formal instability limit \citep[e.g.][]{Raymond_2009}. Furthermore, Figure \ref{fig:ecc-vs-Nce} shows that any additional evolution in these systems is only likely to further increase the orbital eccentricities.

%
%
\section{Discussion}
\label{sec:discussion}

\subsection{Difficulty of the Kozai-Lidov interpretation}
\label{sec:discussion:kozai}

An unfortunate implication of our results is that inferring the presence of the Kozai-Lidov effect is very difficult. The fact that several highly eccentric exoplanets have known wide stellar companions is suggestive \citep[e.g. HD 20782 b, HD 4113 b, HD 80606 b;][]{Desidera_2007,Frith_2013}, but this is merely circumstantial evidence and does not imply causation.

If a much larger sample of highly eccentric exoplanets became available, it might be possible to test correlations between extreme eccentricities and the presence of stellar companions. But that leads into another complication: the Kozai-Lidov mechanism and dynamical instabilities are not mutually exclusive. It is possible for Kozai-Lidov oscillations to \textit{trigger} an instability in a system that would otherwise be stable \citep[e.g.][]{Malmberg_2007}, and it is possible for Kozai-Lidov oscillations to {increase} the eccentricity of a planet after an instability.

\citet{Kaib_2013} showed that the eccentricities of giant exoplanets are statistically higher in systems with wide (>1000 AU) binary companions than in systems with closer binary companions or single stars. This may be explained by variations in wide binary orbits that are driven by Galactic perturbations and which disrupt their planetary systems \citep{Kaib_2013}. This seems to indicate that the Kozai-Lidov effect is probably not central to the high-eccentricity exoplanet population but is instead just a late-stage side effect of other processes.

\subsection{Statistical biases}
\label{sec:discussion:biases}

\citet{Zakamska_2011} investigated potential biases in the measurements of orbital eccentricities from radial velocity data. Unsurprisingly, low signal-to-noise ratio leads to larger uncertainties. In addition, the fact that eccentricities cannot be negative causes a subtle bias toward larger eccentricities for nearly circular orbits. For example, a planet on a perfectly circular orbit may have its eccentricity measured as $e = 0.05$ but it cannot be measured as $e = -0.05$. Both of these biases affect low-eccentricity orbits the most. Since our investigation is focused on the highest eccentricities, we are effectively taking the radial velocity sample with the most reliable eccentricity estimates.

%
%
\subsection{Implications for habitability}
Previous works have shown that strong dynamical instabilities that produce very eccentric giant planets will typically destroy, by ejection or collision with the central star, any terrestrial planets present in the system \citep{Veras_2006,Raymond_2010,Raymond_2011,Carrera_2016}. Giant planets as eccentric as those considered here would certainly wipe out the terrestrial zone. However, previous authors have shown that moons of giant planets can often survive close encounters, with survival rates of tens of percent for close-in moons \citep{Gong_2013,Hong_2018}. Future work should investigate whether giant planet moons can survive the large \textit{number} of close encounters needed to produce a giant-planet eccentricity of $e > 0.95$. While those systems will certainly be quite rare, they would be some of the most dynamically interesting. In addition, an exomoon of an extremely eccentric giant planet is possibly the most extreme environment that may still permit liquid water on the surface, and thus, habitability. Any such moon would very likely experience a prolonged deep-freeze winter with only a brief summer, but previous works suggest that even that type of environment may still be habitable \citep{Williams_2002,Dressing_2010}. For a planet orbiting a Sun-like star, liquid water on the surface can be maintained up to an eccentricity of around $e \sim 0.6$ \citep{Bolmont_2016}.

%
%
\section{Conclusions}
\label{sec:conclusions}

We have modelled the dynamical evolution of unstable planetary systems containing three Jupiter-mass planets. Planet--planet scattering leads to the ejection and collision of some planets leaving others on eccentric orbits. The eccentricity distribution of observed giant ($m \sin(I) > 1 M_{\rm Jup}$) exoplanets with eccentricities above 0.3 is consistent with all of them being the result of planet--planet scattering. Significantly, we find that some planets are left on extremely eccentric orbits ($e > 0.95$). These systems have been missed in earlier work for two reasons: many are on very wide orbits (wider than the semi-major axis cutoff used in many studies); and those closer in have very close periastrons which were beyond the numerical resolution of simulations. Thus planet--planet scattering could be the mechanism responsible for all observed eccentric orbits. The discovery of planets with extremely eccentric orbits does not necessarily signal the action of the Kozai-Lidov mechanism.



%
%
\begin{acknowledgements}
We thank Eric Ford for helpful discussions of observational biases in radial velocity surveys.

D.C.'s research was supported by an appointment to the NASA Postdoctoral Program within NASA's Nexus for Exoplanet System Science (NExSS), administered by Universities Space Research Association under contract with NASA. D.C.\,acknowledges support from NASA Exoplanet Research Program award NNX15AE21G.  The Center for Exoplanets and Habitable Worlds is supported by the Pennsylvania State University, the Eberly College of Science, and the Pennsylvania Space Grant Consortium. This research or portions of this research were conducted with Advanced CyberInfrastructure computational resources provided by The Institute for CyberScience at The Pennsylvania State University (\texttt{http://ics.psu.edu}), including the CyberLAMP cluster supported by NSF grant MRI-1626251.

S.N.R.\,thanks the NASA Astrobiology Institute's Virtual Planetary Laboratory Lead Team, funded under solicitation NNH12ZDA002C and cooperative agreement no. NNA13AA93A.

M.B.D.\,acknowledges the support of the project grant ``IMPACT'' from the Knut and Alice Wallenberg Foundation (KAW 2014.0017).
\end{acknowledgements}

%
%
\bibliographystyle{aa}
\bibliography{references}

%
%
\appendix

\section{Numerical errors}
\label{sec:appendix}

Here we take an in-depth look at the numerical integration errors and their implications.

\begin{figure}[hb!]
        \includegraphics[width=\columnwidth]{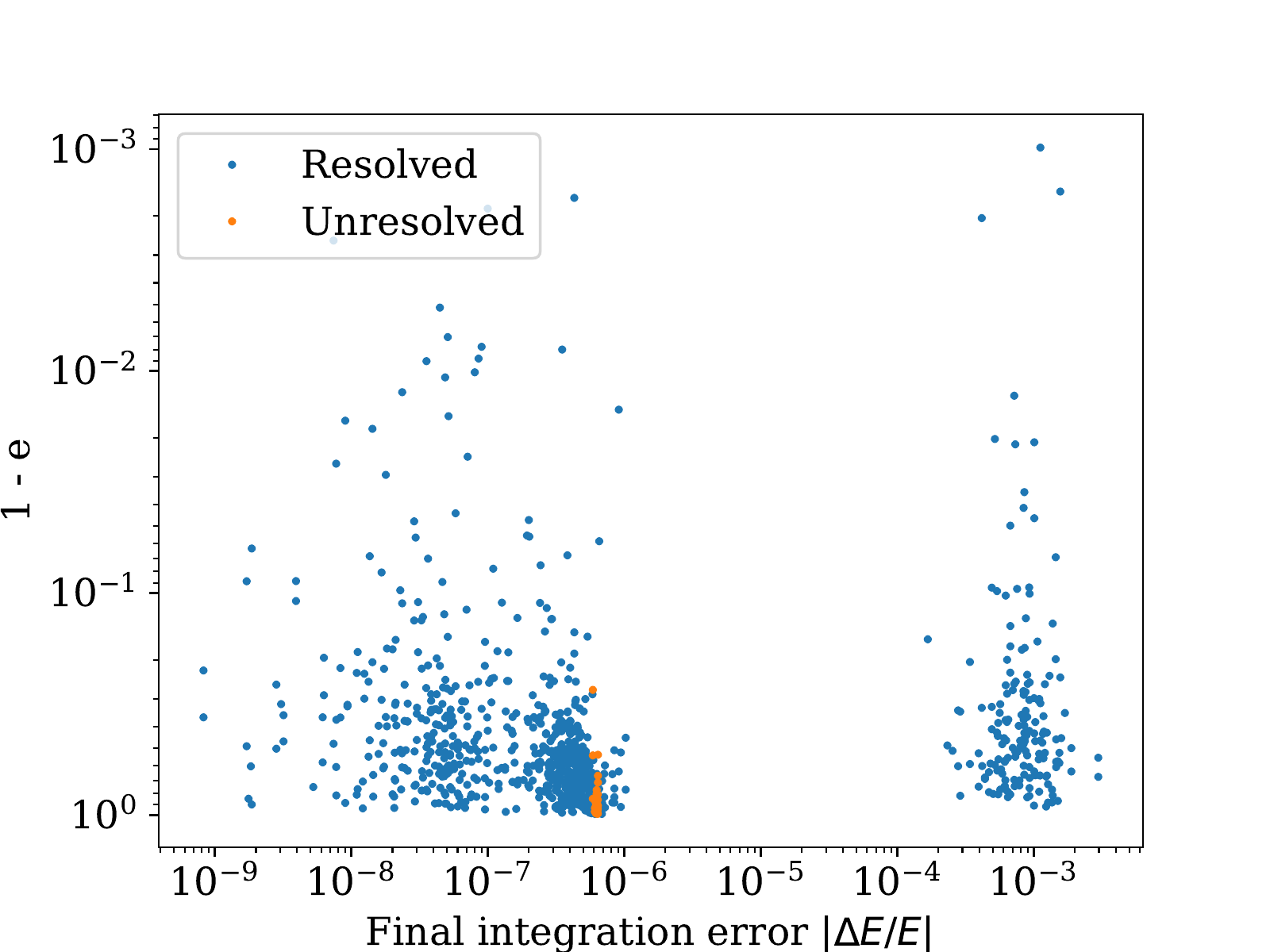}
    \caption{Final eccentricities of all the simulated giant planets against the total accumulated error in energy at the end of the simulation. The lack of correlation between eccentricity and simulation error shows that extreme eccentricities are not a numerical artefact.}
    \label{fig:errors}
\end{figure}

\begin{figure}[hb!]
        \includegraphics[width=\columnwidth]{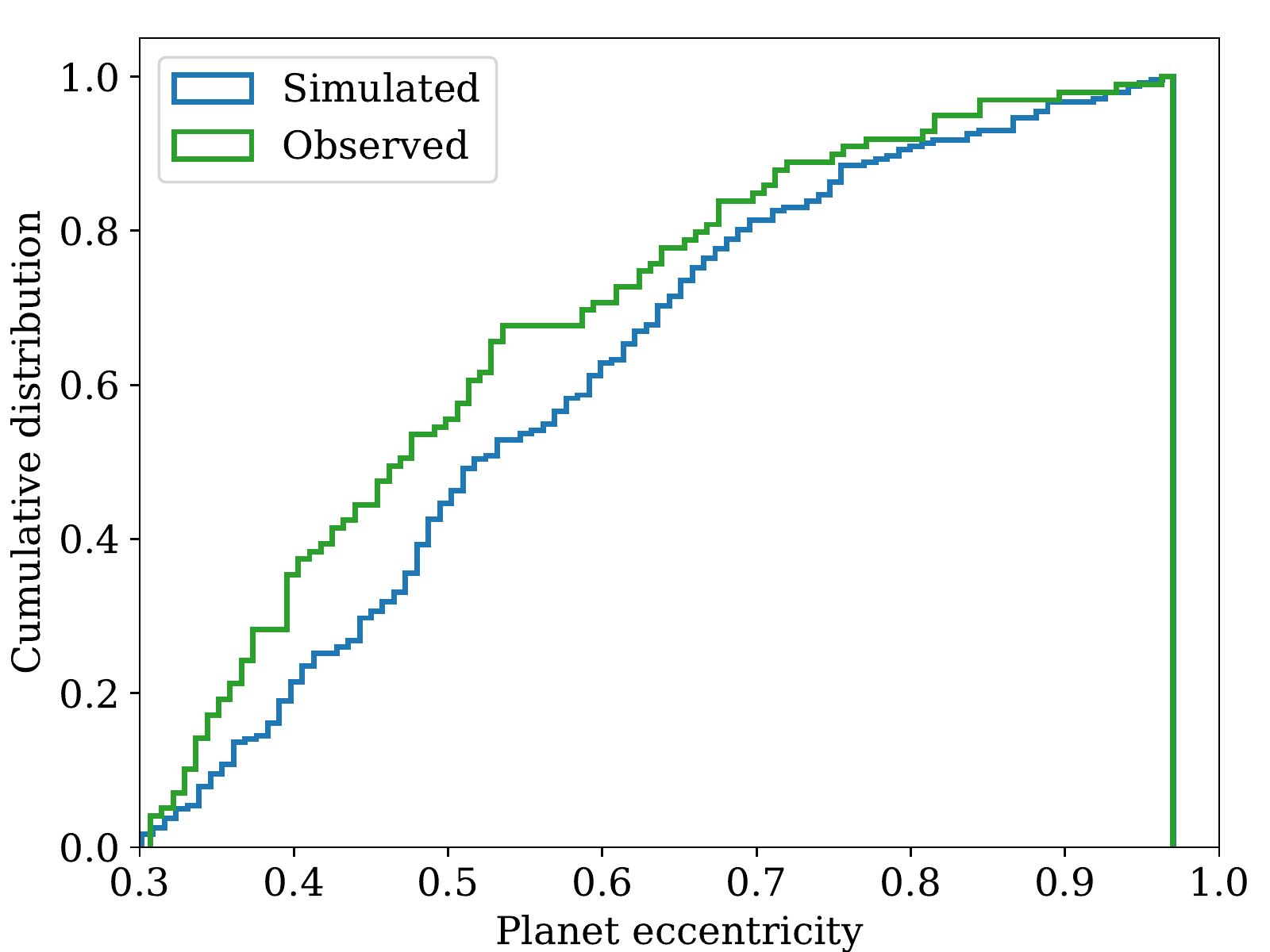}
    \caption{Similar to Figure \ref{fig:sim_vs_obs}, but {excluding} all runs with $|\Delta E/E| > 10^{-5}$. The plot shows the cumulative eccentricity distribution for observed and simulated giant planets ($m \sin(I) > 1 M_{\rm Jup}$) with eccentricity $e > 0.3$. The simulated sample contains only the planets that are resolved, have $a < 5$ AU, and have integration errors $|\Delta E/E| < 10^{-5}$. Compared with Figure \ref{fig:sim_vs_obs}, none of the scientific conclusions change.}
    \label{fig:sim_vs_obs_small_dE}
\end{figure}

Figure \ref{fig:errors} shows the final eccentricities against the total accumulated energy error $|\Delta E/E|$ at the end of the simulation. Clearly there is no broad correlation between final eccentricities and correlation errors. The cluster with $|\Delta E/E| > 10^{-5}$ has a somewhat higher rate of planets with $e > 0.9$ (9.6\% vs. 5.3\%), but because those runs are fewer in number, most planets with $e > 0.9$ come from the cluster with $|\Delta E/E| < 10^{-5}$ (36 vs 16). More importantly, Figure \ref{fig:sim_vs_obs_small_dE} shows what Figure \ref{fig:sim_vs_obs} would look like if all runs with $|\Delta E/E| > 10^{-5}$ were removed. It is clear that none of the scientific conclusions would change between the two figures.

Having said all this, Figure \ref{fig:errors} shows a strong bimodality in $|\Delta E/E|$ that demands an explanation. We determined that the cluster of runs with $|\Delta E/E| > 10^{-5}$ corresponds {exactly} to the runs where at least one planet collided with the central star. When a planet is removed from a simulation,  the energy of that planet is removed as well. The \textsc{mercury} code tries to account for this, but we suspect that it does so imperfectly. In the \textsc{mercury} code, collisions with the central star involve a two-body approximation of the orbit of the planet (i.e. it ignores the other planets). Therefore, it is not surprising that they would incur an error in the order of $m_{\rm pl}/M_\star \sim 10^{-3}$. In fact, since the colliding planet is closer to the star than to the other planets, it is reasonable that the error would typically be slightly lower than $m_{\rm pl}/M_\star \sim 10^{-3}$.

\end{document}